\documentclass{PoS}
\usepackage[utf8]{inputenc}
\usepackage{amsmath}
\usepackage{amsfonts}
\usepackage{textcomp}
\usepackage{graphicx}
\usepackage{dcolumn}
\usepackage{tikz,etoolbox}
\usepackage[point,rounding]{rccol}
\usepackage{booktabs}
\usepackage{varwidth}
\usepackage{colortbl}
\usepackage{color}
\usepackage{bm}
\newcommand{\pTm}{p_{\mbox{\tiny T}}}

\newcommand{\Fig}[1]       {Fig.~\ref{#1}}
\newcommand{\Figure}[1]    {Figure~\ref{#1}}

\newcommand{\rpp}          {\ensuremath{R_{\mathrm{pPb}}}}
\newcommand{\sNN}{\sqrt{s_{\mbox{\tiny NN}}}}

\title{Light meson nuclear modification factor in p-Pb collisions over an unprecedented $\bm{p_\mathrm{T}}$ range with ALICE}

\ShortTitle{$\pi^0$ and $\eta$ $\rpp$ to high $\pTm$  with ALICE}

\author{\speaker{Nicolas Schmidt}\textbf{, on behalf of the ALICE collaboration}\\
Oak Ridge National Laboratory \& IKF Frankfurt\\
E-mail: \email{nicolas.schmidt@cern.ch}}

\abstract{
Light neutral meson differential invariant cross section and nuclear modification factor measurements have been carried out with the ALICE detector at the CERN LHC in pp collisions at $\sqrt{s}=8$~TeV and p--Pb collisions at $\sNN=8.16$~TeV.
The analysis combines results from several partially independent reconstruction techniques where the $\pi^0$ and $\eta$ meson decay photons were detected with the electromagnetic calorimeter, EMCal, the photon spectrometer, PHOS, or via reconstruction of $e^+e^-$ pairs from conversions in the ALICE detector material using the central tracking system.
The neutral pion measurement reaching a $\pTm$ of 200~GeV/$c$ poses as the highest measured identified particle spectrum to date while the $\eta$ meson is measured to an unprecedented $\pTm$ of 50 GeV/$c$. The spectra are found to be generally overestimated by NLO pQCD calculations.
The nuclear modification factors of both mesons exhibit a suppression for $\pTm<10$~GeV/$c$ which is stronger compared to previous measurements at $\sNN=5.02$~TeV and consistent with CGC and cold nuclear matter energy loss calculations. For $\pTm>10$~GeV/$c$, $\rpp$ is consistent with unity and theory predictions.}

\FullConference{%
  HardProbes2020\\
  1-6 June 2020\\
  Austin, Texas}

\begin{document}

    \section{Introduction}
    Light neutral meson measurements in small systems like proton-proton and proton-nucleus collisions at LHC energies over a large transverse momentum range provide important insights into particle production mechanisms and their modifications at small parton fractional momentum $x$ and high parton densities.
    The nuclear modification factor $\rpp$ presents these modifications by deviating from unity and shows distinct features depending on $\pTm$ as seen in previous measurements at RHIC and LHC energies.
    The low $\pTm$ region has been covered by several measurements and is well understood for various particle species where nuclear shadowing and saturation effects are dominant and can be described using nuclear parton distribution functions (nPDFs) \cite{Kovarik:2015cma} and the Color-Glass-Condensate (CGC) framework \cite{Lappi:2013zma}, respectively.
    The high transverse momentum region of identified particle measurements is so far very little explored and the effects of energy loss in cold nuclear matter or anti-shadowing are not yet fully constrained.
    New measurements covering large $\pTm$ ranges can therefore provide additional constraints for theory calculations and allow one to disentangle initial and final state effects in the modification of particle production.

    \section{Detector description and datasets}
    The central barrel of ALICE \cite{Abelev:2014ffa} provides multiple detector systems for particle tracking, identification and reconstruction.
    The presented neutral meson measurements require the reconstruction of decay photons for which ALICE provides two calorimeter-based methods using the Electromagnetic Calorimeter (EMCal) and the Photon Spectrometer (PHOS) as well as the reconstruction of photons from the e$^+$e$^-$ pairs produced by conversions in the detector material using the Inner Tracking System (ITS) and the Time Projection Chamber (TPC).
    The data used in the presented measurements is from pp ($\sqrt{s}=8$~TeV) and p--Pb ($\sNN=8.16$~TeV) collisions recorded in 2012 and 2016, respectively. 
    Both systems have slightly different detector configurations due to upgrades installed during the LHC long shutdown 1 which are mentioned in the following.
    The EMCal is a lead-scintillator sampling calorimeter that covered $\Delta\varphi=100^\circ$ for $|\eta|<0.7$ in 2012 and was extended by the DCal in 2015 adding an additional $\Delta\varphi=68^\circ$ for $0.22<|\eta|<0.7$ opposite in azimuth. 
    Together they consist of 18240 cells with $\Delta\eta\times\Delta\varphi=0.0143\times0.0143$ rad$^2$ in size which provide an energy resolution of $\sigma_E/E=4.8\%/E\oplus11.3\%/\sqrt{E}\oplus1.7\%$ with $E$ in units of GeV.
    The PHOS is a lead tungstate crystal calorimeter that covered in its 2012 configuration $\Delta\varphi=60^\circ$ for $|\eta|<0.13$ and is made of cells with $22\times22$ mm$^2$ which are close to the Moli\`ere radius of 2 cm that provide an excellent resolution of $\sigma_E/E=1.8\%/E\oplus3.3\%/\sqrt{E}\oplus1.1\%$.
    The ITS is a six layered cylindrical silicon detector system with two layers each of Silicon Pixel (SPD), Silicon Drift (SDD) and Silicon Strip (SSD) detectors used for the reconstruction of primary and secondary vertices. 
    The TPC is a large cylindrical drift chamber that covers $|\eta|<0.9$ and provides together with the ITS particle tracking as well as particle identification via energy loss signals d$E$/d$x$ down to $\pTm\approx100$ and 50 MeV/$c$ for primary and secondary tracks, respectively.
    The V0 detectors are two scintillator arrays covering the full azimuth and the forward pseudorapidities $2.8<\eta<5.1$ and $-3.7<\eta<-1.7$.
    They are used to define the minimum bias trigger which requires a coincidence signal in both arrays and for determining the event multiplicity.
    Further event triggers are provided by both calorimeters based on large energy deposits in small arrays of cells.
    
    \section{Photon and meson reconstruction in ALICE}
    The decay photons of the $\pi^0$ and $\eta$ mesons are reconstructed with three reconstruction techniques.
    The photon conversion method (PCM) uses the ITS and TPC to reconstruct photons from e$^+$e$^-$ pairs produced in photon conversions that occur with an approximate 8.9\% probability in the inner detector material of ALICE.
    The identification criteria are similar to those given in Refs. \cite{Acharya:2018hzf,Acharya:2017tlv} with additional improvements to increase the reconstruction efficiency at high $\pTm$ by loosening the constraints on the variable $q_\mathrm{T}$ which is the projection of the electron momentum to the photon candidate momentum.
    The photon reconstruction with PHOS and EMCal is based on the clusterization of energy deposits in the calorimeter cells from electromagnetic showers induced by photons or electrons. The selection criteria for the calorimetric analyses are adapted from Refs. \cite{Acharya:2018hzf,Acharya:2017tlv} for the EMCal while the PHOS has only been used in the pp reference measurement.
    For EMCal additional improvements were made regarding the nonlinearity of the calorimeter response at high energies. The corresponding studies resulted in two nonlinearity corrections which account for a hardware nonlinearity of up to 14\% at cluster energies of 200~GeV and a residual energy and position correction which is applied in simulations to match the $\pi^0$ invariant mass peak positions in data resulting in a per-mille level of agreement.
    
    The neutral mesons are reconstructed using an invariant mass technique where photons reconstructed with the same technique are combined, called PCM, EMC or PHOS, or following a hybrid approach where a PCM photon is paired with an EMCal photon called PCM-EMC.
    The latter profits from the high resolution of the conversion photon and the high efficiency of the EMC method and is able to cover a large transverse momentum range while the pure PCM method is strongest at low $\pTm$ and the pure EMC method suffers, despite its high efficiency, from cluster merging starting from 16~GeV/$c$ for the $\pi^0$.
    Combinatorial background in this method is removed using an event mixing technique and raw yields are obtained by integration of wide ranges around the peak position which is obtained from a combined gaussian and one-sided exponential fit on the invariant mass distributions.
    
    In addition to the techniques based on the invariant mass, a particle identification method based on merged photon clusters from $\pi^0$ decays is employed, called mEMC \cite{Acharya:2017hyu}.
    This method exploits that the opening angle of high momentum ($\pTm>16$~GeV/$c$) $\pi^0$ mesons are too small to induce two separate showers in the EMCal.
    Instead the showers overlap and thus produce a single shower of ``elliptical'' shape which can be distinguished from single photon clusters via the shower shape parameter $\sigma_\mathrm{long}^2$ which can be interpreted as the long axis of the shower ellipse. 
    Photon clusters are located at small values of $\sigma_\mathrm{long}^2$ thus the application of a selection with $\sigma_\mathrm{long}^2>0.2$ allows us to reject single photon background resulting in a purity between 81--87\% depending on $\pTm$.
    As this method relies on a good description of $\sigma_\mathrm{long}^2$ in simulation compared to data, an additional detector effect, called cross-talk, had to be emulated within the same readout card as described in \cite{Acharya:2019jkx}.

         \begin{figure}[t]
    \center
    \includegraphics[width=.8\textwidth]{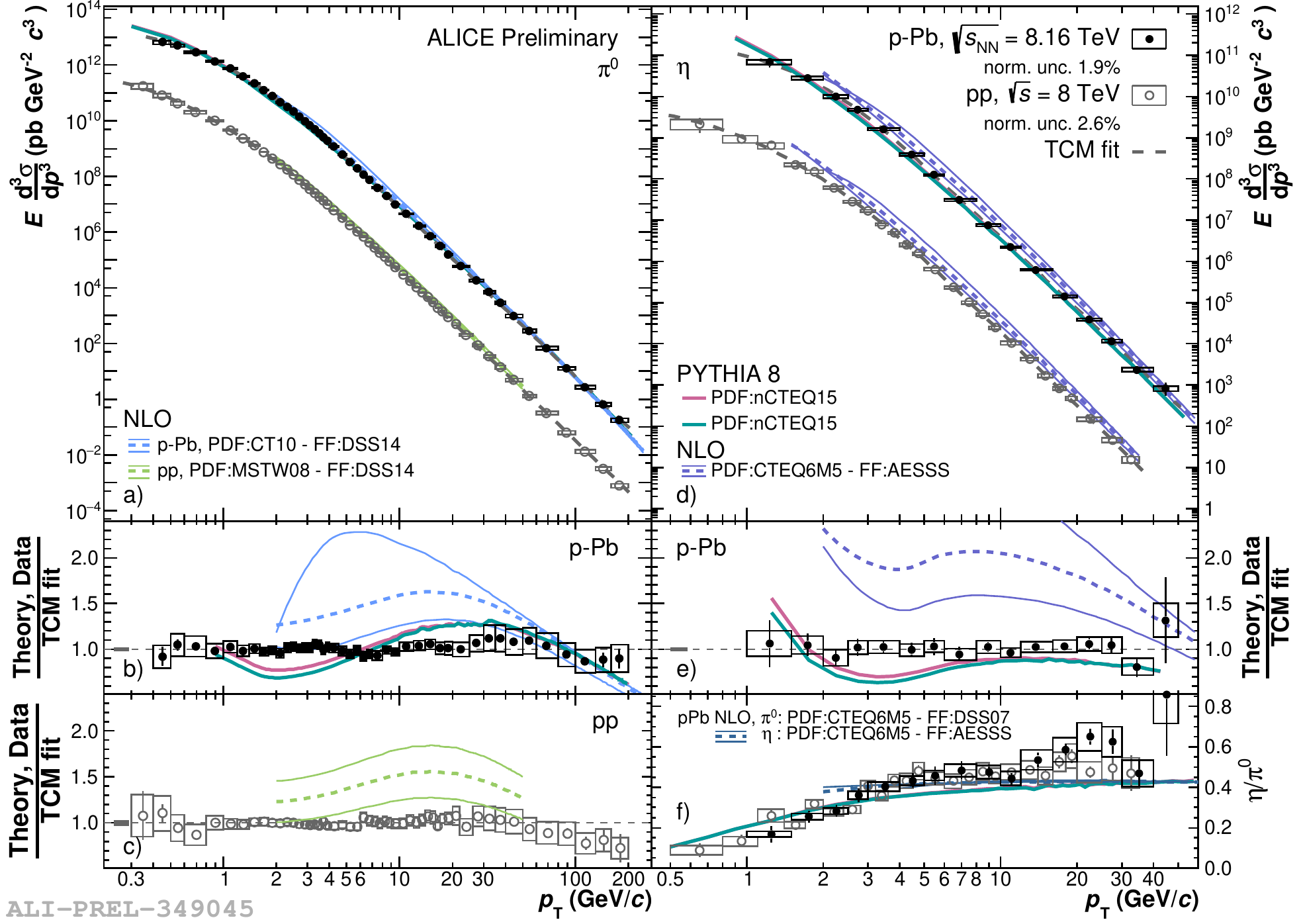}
      \caption{Top: Differential invariant cross section spectra of $\pi^0$ and $\eta$ mesons in pp collisions at $\sqrt{s}=8$~TeV and p--Pb collisions at $\sNN=8.16$~TeV together with two-component model fits and Pythia 8 calculations with different nPDFs \cite{Kovarik:2015cma} as well as NLO calculations using CT10, MSTW08 and CTEQ6M5 PDFs with DSS14 and AESSS fragmentation functions. Lower panels: Ratio of the p--Pb $\sNN=8.16$~TeV $\pi^0$ and $\eta$ meson spectra and their corresponding theory calculations to their respective two-component model fit. Statistical uncertainties are given by the vertical lines while systematic uncertainties are shown as boxes. Bottom right: $\eta/\pi^0$ ratio in both collision systems together with NLO calculations.}
    \label{fig:spectra}
  \end{figure}
    
    \section{Spectra and nuclear modification measurements}
    The $\pi^0$ and $\eta$ meson differential invariant cross sections have been measured in pp collisions at $\sqrt{s}=8$~TeV \cite{Acharya:2017tlv} and p--Pb collisions at $\sNN=8.16$~TeV with several partially independent reconstruction techniques. 
    The results of the individual measurements were combined using the Best Linear Unbiased Estimates (BLUE) method accounting for correlations of uncertainties. The p--Pb measurement is therefore a combination of the PCM, EMC, PCM-EMC and mEMC methods while for the pp measurement also PHOS was used to reconstruct the $\pi^0$. 
    The spectra together with two-component model (TCM) fits and NLO \cite{Kovarik:2015cma} as well as Pythia 8 calculations are shown in \Figure{fig:spectra}. In addition, the ratios of the spectra and theory predictions to the TCM fits are given which show an overestimation of the spectra by NLO calculations especially for the $\eta$ meson.
    The $\eta/\pi^0$ ratio is presented in the same figure (bottom right) and found to be consistent between both collision systems with a constant value of about $0.49$ for $\pTm>4$~GeV/$c$ in the p--Pb measurement.
    
\begin{figure}[t]
        \center
        \includegraphics[width=.48\textwidth]{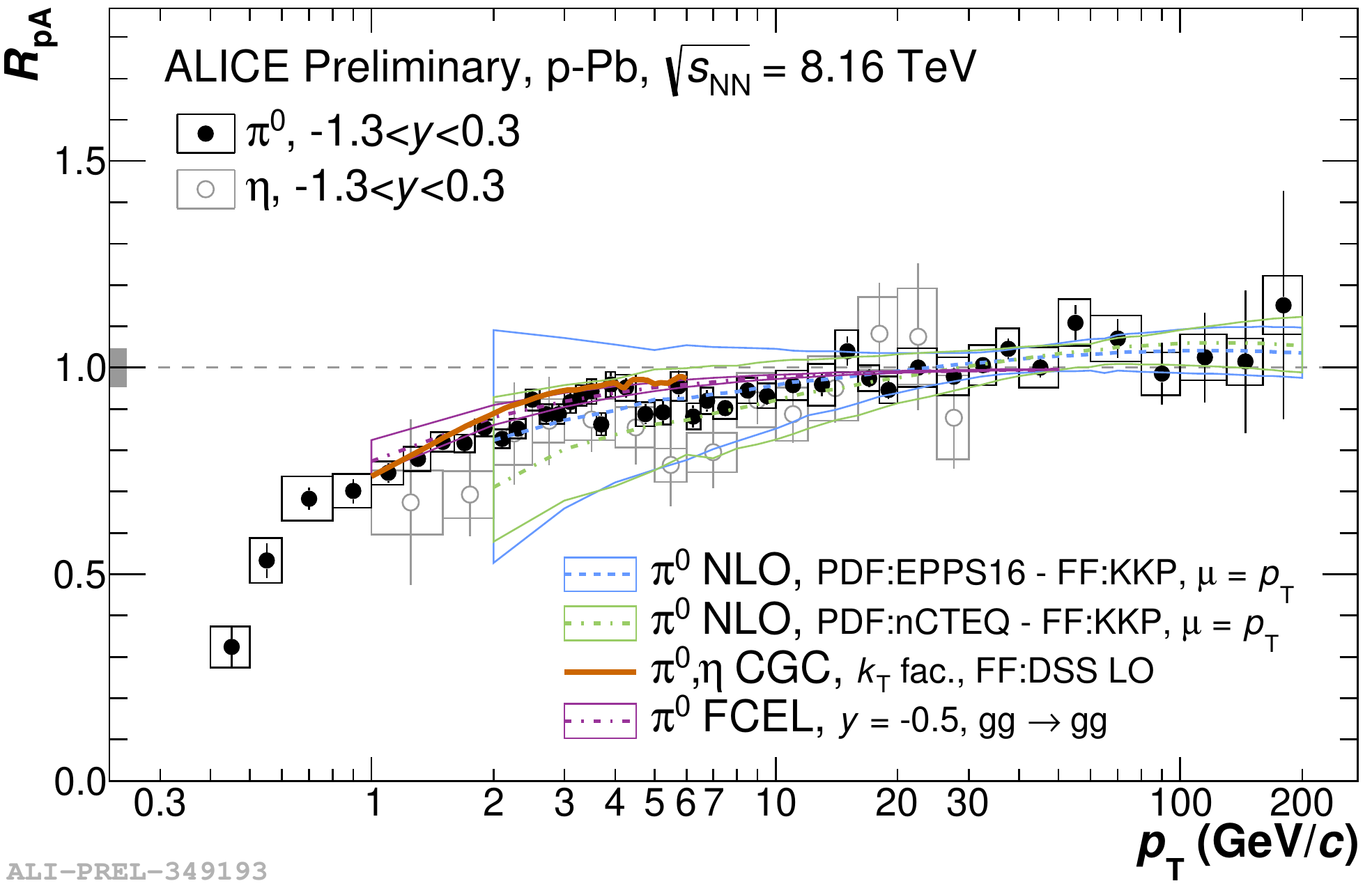}\hspace{0.2cm}
        \includegraphics[width=.48\textwidth]{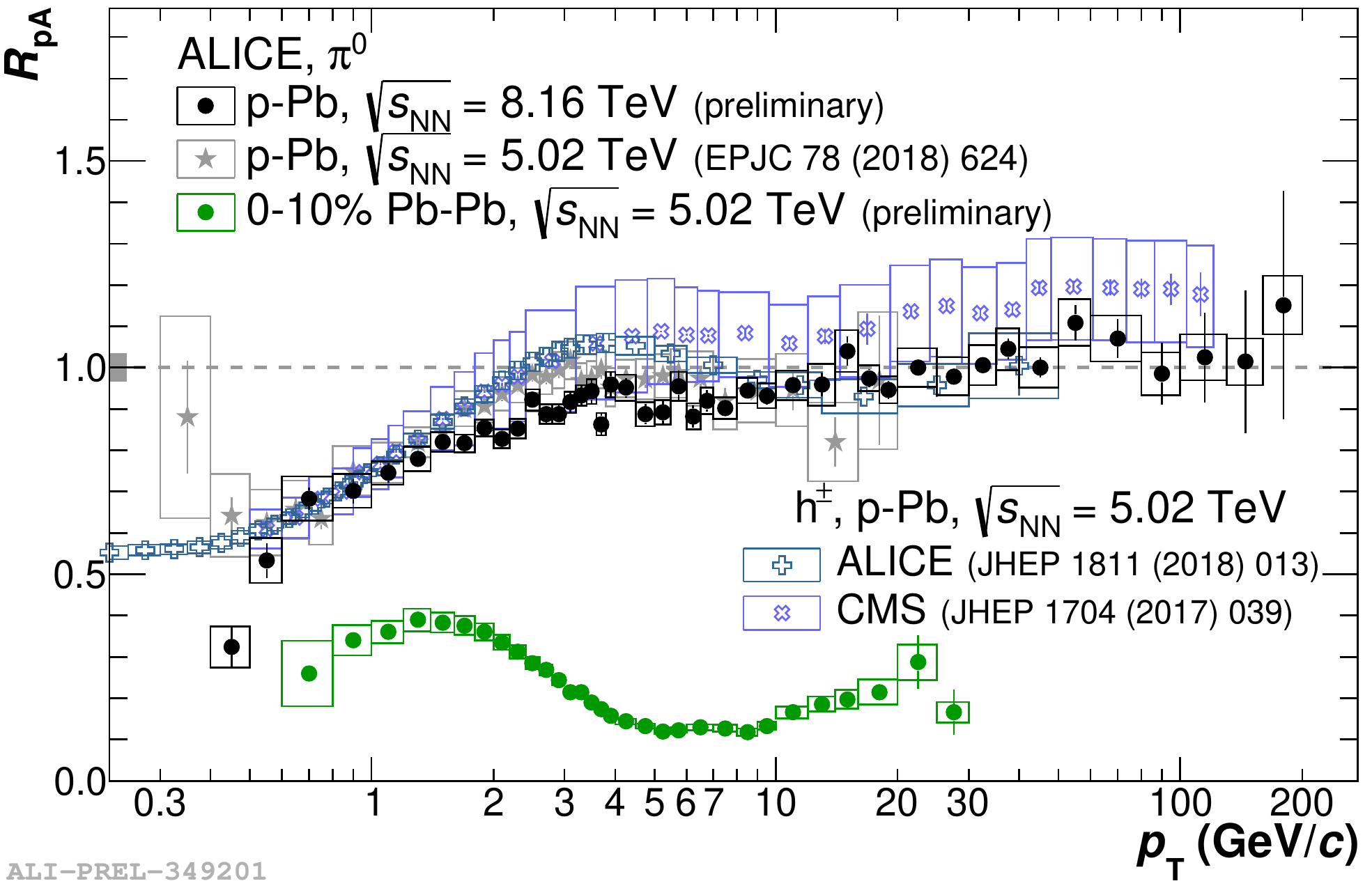}
        \caption{Left: Nuclear modification factor of $\pi^0$ and $\eta$ mesons in p--Pb collisions at $\sNN=8.16$~TeV together with two NLO calculations as well as calculations from the Color-Glass-Condensate (CGC) framework \cite{Lappi:2013zma} and cold nuclear matter energy loss calculations (FCEL) \cite{Arleo:2020hat}. Right: Nuclear modification factors of neutral pions in p--Pb collisions at $\sNN=5.02$ and 8.16~TeV and central Pb--Pb collisions at $\sNN=5.02$~TeV together with charged hadron measurements from ALICE and CMS in p--Pb collisions at $\sNN=5.02$~TeV. Statistical uncertainties are given by the vertical lines while systematic uncertainties are shown as boxes.}
        \label{fig:rppb}
    \end{figure}

    The nuclear modification factors for both mesons were calculated using the measured spectra in pp and p--Pb with $\rpp = \sigma_{\mathrm{pPb}}^{\pi^0,\eta} / (208\cdot\sigma_{\mathrm{pp}}^{\pi^0,\eta})$ where the reference measurement was scaled to the p--Pb collision energy as well as corrected for the rapidity shift using Pythia8 Monash2013.
    The resulting $\rpp$ are shown in \Fig{fig:rppb} (left) for both mesons together with NLO calculations using different nPDFs \cite{Kovarik:2015cma}. In addition, calculations from the CGC framework \cite{Lappi:2013zma} and cold nuclear matter energy loss (FCEL) \cite{Arleo:2020hat} are able to describe the $\pTm<10$~GeV/$c$ region.
    Furthermore, a comparison to a previous measurement in p--Pb $\sNN=5.02$~TeV, given in \Fig{fig:rppb} (right) shows a $7\%$ stronger suppression for $\pTm<6$~GeV/$c$ in the new data which is supported by the CGC calculations hinting at gluon saturation effects. The neutral meson $\rpp$ are found to be consistent with unity for $\pTm>10$~GeV/$c$ while the moderate enhancement of the CMS charged hadron measurement which was interpreted as an anti-shadowing effect shown in the same figure is not observed in the presented measurement.
  
    \section{Conclusions}
    The differential invariant cross sections and nuclear modification factors of $\pi^0$ and $\eta$ mesons have been measured in p--Pb collisions at $\sNN=8.16$~TeV over unprecedented transverse momentum ranges of $0.4<\pTm<200$~GeV/$c$ and $1<\pTm<50$~GeV/$c$, respectively. The reference $\pi^0$ measurement has been improved and extended up to the same $\pTm$ using the merged cluster analysis in order to calculate $\rpp$.
    The nuclear modification factors were found to be consistent with CGC and cold nuclear matter energy loss calculations at low $\pTm$ with a stronger suppression compared to previous measurements at a lower center-of-mass energy.
    For high transverse momenta of $\pTm>10$~GeV/$c$, $\rpp$ is found to be consistent with unity and theory predictions. The moderate enhancement observed of the CMS charged hadron measurement is not observed.


\begin{thebibliography}{99}
  
  \bibitem{Kovarik:2015cma}
  Kovarik \textit{et al.}, Phys.\ Rev.\ D {\bf 93}, 085037 (2016)
  \bibitem{Lappi:2013zma}
  Lappi and Mäntysaari, Phys.\ Rev.\ D {\bf 88}, 114020 (2013)
  \bibitem{Abelev:2014ffa} 
  Abelev \textit{et al.}, \textbf{ALICE} Collaboration, \textit{Int. J. Mod. Phys.} \textbf{A29} (2014) 1430044
  \bibitem{Acharya:2018hzf}
  Acharya \textit{et al.}, \textbf{ALICE} Collaboration, EPJC {\bf 78}, 8 (2018) 624
  \bibitem{Acharya:2017tlv}
  Acharya \textit{et al.}, \textbf{ALICE} Collaboration, EPJC {\bf 78}, 3 (2018) 263
  \bibitem{Acharya:2019jkx}
  Acharya \textit{et al.}, \textbf{ALICE} Collaboration, EPJC {\bf 79}, 11 (2019) 896
  \bibitem{Acharya:2017hyu}
  Acharya \textit{et al.}, \textbf{ALICE} Collaboration, EPJC {\bf 77}, 9 (2017) 586
  \bibitem{Arleo:2020hat}
  Arleo \textit{et al.}, arXiv:2003.06337 [hep-ph]
  
  
  \end{thebibliography}
\end{document}